\documentclass[aip,jcp,amsmath,amssymb,amsfonts,reprint,groupedaddress,floatfix]{revtex4-1}
\usepackage{graphicx}
\usepackage{dcolumn} 
\usepackage{ifthen}
\usepackage{xcolor}
\newcommand{\nn}[0]{\nonumber \\}

\ifthenelse{\isundefined{\vr}}{ \newcommand{\vr}[0]{ {\boldsymbol{r}} } }{ \renewcommand{\vr}[0]{ {\boldsymbol{r}} } }

\newcommand{\be}{\begin{equation}}
\newcommand{\ee}{\end{equation}}
\newcommand{\bea}{\begin{eqnarray}}
\newcommand{\eea}{\end{eqnarray}}

\newcommand{\bra}[0]{\langle}
\newcommand{\ket}[0]{\rangle}

\newcommand{\xc}[0]{\mathrm{xc}}

\def\b{\beta}

\def\G{\Gamma}
\def\d{\delta}

\def\l{\lambda}

\def\s{\sigma}
\def\S{\Sigma}

\def\w{\omega}

\def\bra{\langle}
\def\ket{\rangle}

\def\xc{{\rm xc}}
\def\x{{\rm x}}

\begin{document}

\title{Strengths and limitations of the adiabatic exact-exchange kernel for total energy calculations}

\author{Maria Hellgren}
\email[]{maria.hellgren@upmc.fr}
\affiliation{Sorbonne Universit\'e, MNHN, UMR CNRS 7590, IMPMC, 4 place Jussieu, 75005 Paris, France}

\author{Lucas Baguet} 
\affiliation{ CEA, DAM, DIF, F-91297 Arpajon, France and Universit\'e Paris-Saclay, CEA, Laboratoire Mati\`ere en Conditions Extr\^emes, 91680 Bruy\`eres-le-Ch\^atel, France}

\begin{abstract}
We investigate the adiabatic approximation to the exact-exchange kernel for calculating correlation energies within the adiabatic-connection fluctuation-dissipation framework of time-dependent density functional theory. A numerical study is performed on a set of systems having bonds of different character (H$_2$ and N$_2$ molecules, H-chain, H$_2$-dimer, solid-Ar and the H$_2$O-dimer).  We find that the adiabatic kernel can be sufficient in strongly bound covalent systems, yielding similar bond lengths and binding energies. However, for non-covalent systems the adiabatic kernel introduces significant errors around equilibrium geometry, systematically overestimating the interaction energy. 
The origin of this behaviour is investigated by studying a model dimer composed of one-dimensional closed-shell atoms interacting via soft-Coulomb potentials. The kernel is shown to exhibit a strong frequency dependence at small to intermediate atomic separation that affects both the low-energy spectrum and the exchange-correlation hole obtained from the corresponding diagonal of the two-particle density matrix.
\end{abstract}

\date{\today}
\maketitle
\section{Introduction} \label{SEC:Introduction}
In the continuous effort to improve the accuracy of the exchange-correlation (xc) energy in density functional theory (DFT),\cite{PhysRev.136.B864,PhysRev.140.A1133,parryang,doi:10.1063/1.1390175,RappoportBurke:09} time-dependent (TD) DFT\cite{PhysRevLett.52.997} has emerged as a promising route for advanced functional constructions. TDDFT gives access to the dynamical linear density response function which not only contain the spectrum but also provides the exact correlation energy via the adiabatic-connection fluctuation dissipation (ACFD) framework.\cite{LANGRETH19751425,PhysRevB.15.2884,tddftbook,PhysRevB.61.13431,PhysRevB.105.035123,perdew2021interpretations,PhysRevB.65.235109,doi:10.1063/1.1884112,doi:10.1080/00268976.2011.614282} 

Within linear response TDDFT the central quantity to approximate is the so-called xc kernel, defined as the functional derivative of the TD xc potential with respect to the density.\cite{PhysRevLett.55.2850} Adopting local and semi-local approximations from ground-state DFT generates adiabatic approximations, i.e., the xc potential depends only on the instantaneous density which leads to frequency independent xc kernels. Non-adiabatic approximations have been difficult 
to construct, but several works have shown that they are of crucial importance. For real time simulations\cite{PhysRevLett.119.263401,C8CP03957G} as well as for capturing important features of the spectrum. The latter includes double excitations,\cite{doi:10.1063/1.1651060} inner-shell\cite{doi:10.1063/1.3179756} and charge transfer (CT) excitations,\cite{doi:10.1063/1.1590951,doi:10.1063/1.2387951,doi:10.1021/ja8087482,PhysRevA.85.022514,PhysRevA.88.052507} all requiring an xc kernel with a strong frequency dependence. 

On the other hand, for the ACFD correlation energy the frequency dependence is expected to be less critical since the linear density response function is integrated over the frequency. The problems described above could, therefore, be quantitatively small or cancel out. Indeed, the exact adiabatic kernel of one-dimensional model systems has been shown to capture most of the correlation energy.\cite{PhysRevB.104.125126} It has also been shown that the correlation energies of atoms,\cite{hellgren_vonbarth2010,PhysRevA.73.010502} small molecules\cite{doi:10.1080/00268970903476662} and the electron gas\cite{PhysRevB.61.13431} are relatively insensitive to the adiabatic approximation. Furthermore, improved results for solids have been obtained with adiabatic kernels.\cite{PhysRevLett.112.203001,olsen2019beyond}

In this work we study a larger set of systems and reexamine the role of the frequency dependence of the xc kernel for calculating correlation energies. We focus on the frequency dependent exact-exchange (EXX) kernel\cite{PhysRevA.57.3433} and the corresponding TDEXX response function.\cite{PhysRevB.78.115107} Inserted into the ACFD expression this approximation generates the random phase approximation with exchange (RPAx), shown to be highly accurate on molecules\cite{hellgren_vonbarth2010,doi:10.1080/00268970903476662,doi:10.1063/1.4922517,hellgren_correlation_2018} and solids.\cite{PhysRevResearch.3.033263} An important motivation is to investigate whether the adiabatic EXX kernel can be used for lowering the computational cost of RPAx calculations. Evaluating the kernel at $\w=0$ only, has the potential to reduce the cost by a factor of 8 or more. Another motivation is to gain new theoretical insights that may be useful for the development of approximate xc kernels. 

The paper is organized as follows. In Sec. II we start by briefly reviewing the theory that will be used throughout the paper. In Sec. III we present a numerical study of the RPAx employing the adiabatic EXX kernel on a set of selected systems (H-chain, H$_2$-dimer, N$_2$-molecule, solid-Ar and the H$_2$O-dimer). This is followed by an in-depth analysis of the EXX kernel on a model one-dimensional dimer with soft-Coulomb interactions. 
Finally, in Sec. IV we provide our conclusions. 

\section{Theory}
We start by reviewing the basic theory that will be used in this work. The interaction energy is first written 
in terms of the two-particle density matrix, which is then related to the ACFD expression, in which the interaction energy is written in terms of the linear density response function.
More detailed derivations can be found in, e.g., Refs.~\onlinecite{parryang,baerends1997quantum,vonbarthdft,fw,tddftbook}.

We consider the many-electron Hamiltonian   
\be
\hat H=\hat T+\hat V_{\rm ext}+\hat W,
\ee
where $\hat T$ is the kinetic energy operator, $\hat V_{\rm ext}$ is the external nuclear 
potential operator and $\hat W$ is the electron-electron interaction operator. 
Taking the expectation value of $\hat H$ with respect to the ground-state wave function $\Psi$ we can write the ground-state energy as
\be
E=\bra \hat H\ket=T+E_{\rm ext}+U,
\ee
where the kinetic energy 
\be
T=\bra\hat T\ket= -\frac{1}{2} \int \!d\vr \,[\nabla_{\vr}^2\G^{(1)}(\vr,\vr')]_{\vr'=\vr}
\ee
is given in terms of the spin-summed  
\be
\G^{(1)}(\vr,\vr')=\sum_{\s\s'}\G_{\s\s'}^{(1)}(\vr,\vr')
\ee
one-particle density matrix
\be
\G_{\s\s'}^{(1)}(\vr,\vr')=N\int\!dx_{2-N}\Psi(x,\ldots,x_N)\Psi^*(x',\ldots,x_N),
\ee
($x=(\vr,\s)$ is a composite variable of space and spin coordinates and $N$ is the number of electrons), 
the external potential energy 
\be
E_{\rm ext}=\bra\hat V_{\rm ext}\ket= \int \!d\vr \,n(\vr)v_{\rm ext}(\vr)
\ee
is given in terms of the electron density
\be
n(\vr)=\sum_\s\G_{\s\s}^{(1)}(\vr,\vr)
\ee
and the interaction energy
\be
U=\bra \hat W \ket=\frac{1}{2}\int \!d\vr d\vr'\,v(\vr-\vr')\G^{(2)}(\vr,\vr')
\label{Ugamma}
\ee
in terms of the spin-summed 
\be
\G^{(2)}(\vr,\vr')=\sum_{\s\s'}\G_{\s\s'}^{(2)}(\vr,\vr')
\ee
diagonal of the two-particle density matrix
\be
\G_{\s\s'}^{(2)}(\vr,\vr')=N(N-1)\int \!dx_{3-N}|\Psi(x,x',\ldots,x_N)|^2.
\label{defg2}
\ee
The electron density $n$ describes the probability of finding any of the electrons at $\vr$, and 
$\G^{(2)}$ describes the probability of finding any two electrons at $\vr$ and $\vr'$. 

Although the ground-state energy only depends on the spin-summed density matrices it is interesting to examine their spin structure. From the diagonal of the two-particle density matrix we can determine the conditional probability of finding an electron at $\vr'$ with spin $\s'$ given there is another electron at $\vr$ with spin $\s$ 
\be
n^{{\rm cond}}_{\s'|\s}(\vr'|\vr)=\frac{\G_{\s\s'}^{(2)}(\vr,\vr')}{n_\s(\vr)}.
\ee
This quantity can be split into two terms 
\be
n^{{\rm cond}}_{\s'|\s}(\vr'|\vr)=n_{\sigma'}(\vr')+n^{{\rm hole}}_{\s'|\s}(\vr'|\vr),
\label{ncond}
\ee
where the first term corresponds to independent electrons, i.e., the probability of finding an electron with a 
certain position and spin is completely independent of the position and spin of the other. 
This term, when inserted into Eq.~(\ref{Ugamma}), gives rise to the classical Hartree energy
\be
U_{0}=\frac{1}{2}\int \!d\vr d\vr'\,n(\vr)v(\vr-\vr')n(\vr').
\ee
The second term in Eq.~(\ref{ncond}) is the so-called xc hole. Due to the antisymmetry of the 
wave function the xc hole must be exactly equal to $-n_{\s'}(\vr')$ when two electrons of the same spin $\s'$ are at the same position $\vr'$. Furthermore, Eq.~(\ref{defg2}) leads to a sum-rule which states that the same-spin xc 
hole must integrate to -1, and that of opposite spins to zero.\cite{vonbarthdft} 

The remaining part of the interaction energy, the xc energy ($U_\xc= U-U_0$), can now be written in terms of the xc hole
\be
U_\xc =\frac{1}{2}\sum_{\s\s'}\int \!d\vr d\vr'\,n_{\sigma}(\vr)v(\vr-\vr')n^{\rm hole}_{\s'|\s}(\vr'|\vr).
\label{uxcdef}
\ee
We will use this expression for our analysis in Sec.~III~E.

Let us now derive a second expression for the xc energy. We write the diagonal of the two-particle density 
matrix in second quantization 
\be
\G_{\s\s'}^{(2)}(\vr,\vr')=\bra\Psi|\hat\psi^\dagger_{\s}(\vr)\hat\psi^\dagger_{\s'}(\vr')\hat\psi_{\s'}(\vr')\hat\psi_{\s}(\vr)|\Psi\ket.
\ee 
Using the anticommutation relations of the fermionic field operators $\psi(\psi^\dagger)$ we can write\cite{fw}
\bea
\G_{\s\s'}^{(2)}(\vr,\vr')&=&\bra\Psi|\tilde n_\s(\vr)\tilde n_{\s'}(\vr')|\Psi\ket+ n_\s(\vr) n_{\s'}(\vr')\nn
&&-n_\s(\vr)\d_{\s\s'}\d(\vr-\vr')
\eea
where $\tilde n_\s(\vr)=\hat n_\s(\vr)-n_\s(\vr)$ is the deviation operator. Via the fluctuation dissipation (FD) theorem 
the first term can be shown to be equal to the frequency integrated dynamical linear density response function $\chi$.\cite{LANGRETH19751425,PhysRevB.15.2884} 
The xc energy can thus be written as
\bea
U_\xc&=&-\frac{1}{2}\sum_{\s\s'}\int \!d\vr d\vr'\,v(\vr-\vr')\nn
&&\!\!\!\!\!\!\!\!\!\!\!\!\!\!\!\!\!\!\!\!\!\!\times\left[\int_0^\infty \!\frac{d\w}{\pi} \chi^{\s\s'}(\vr,\vr',i\w)+n_\s(\vr)\d_{\s\s'}\d(\vr-\vr')\right].
\label{uxc}
\eea
This expression allows one to generate approximations to the interaction energy via approximations to $\chi$, and every approximation to $\chi$ corresponds to an approximate xc hole. The sum-rule is trivially fulfilled 
for any physical response function giving zero response to a constant potential. 

There are various ways to approximate the linear density response function. One can proceed via time-dependent Hartree-Fock,\cite{PhysRevLett.102.096404} the Bethe-Salpeter equation\cite{10.21468/SciPostPhys.8.2.020,doi:10.1063/1.4871875} or via linear response TDDFT.\cite{tddftbook} 
In the latter case, the density is obtained from the time-dependent Kohn-Sham (KS) system, defined by an effective one-particle Hamiltonian that gives the exact density via a local xc potential. The interacting linear density response function is then given by 
\be
 {\mbox{\boldmath $\chi$}}= {\mbox{\boldmath $\chi_s$}}+ {\mbox{\boldmath $\chi_s$}} {\mbox{\boldmath $K$}} {\mbox{\boldmath $\chi$}}
\label{tddftLR}
\ee
where
\[
  {\mbox{\boldmath $\chi$}} =
  \left[ {\begin{array}{cc}
    \chi^{\uparrow\uparrow} & \chi^{\downarrow\uparrow} \\
    \chi^{\uparrow\downarrow} & \chi^{\downarrow\downarrow} \\
  \end{array} } \right],
 \,\,\,\,{\mbox{\boldmath $\chi_s$}} =
  \left[ {\begin{array}{cc}
    \chi_s^{\uparrow\uparrow} & \chi_s^{\downarrow\uparrow} \\
    \chi_s^{\uparrow\downarrow} & \chi_s^{\downarrow\downarrow} \\
  \end{array} } \right]
\]
and
\[
  {\mbox{\boldmath $K$}} =
  \left[ {\begin{array}{cc}
    v+f_\xc^{\uparrow\uparrow} &  v+ f_\xc^{\downarrow\uparrow}  \\
     v+  f_\xc^{\uparrow\downarrow} & v+ f_\xc^{\downarrow\downarrow} \\
  \end{array} } \right],
\]
and $\chi^{\s\s'}_s$ is the independent-particle KS response function. The xc kernel $f^{\s\s'}_\xc$ is defined as the functional derivative of the time-dependent xc potential with respect to the density, evaluated at the ground-state density
\be
f^{\s\s'}_\xc(\vr t,\vr't')=\left.\frac{\d v^{\s}_\xc(\vr t)}{\d n_{\s'}(\vr' t')}\right|_{n_{0}}.
\label{fxceq}
\ee
\begin{figure*}[t]
\includegraphics[scale=0.5]{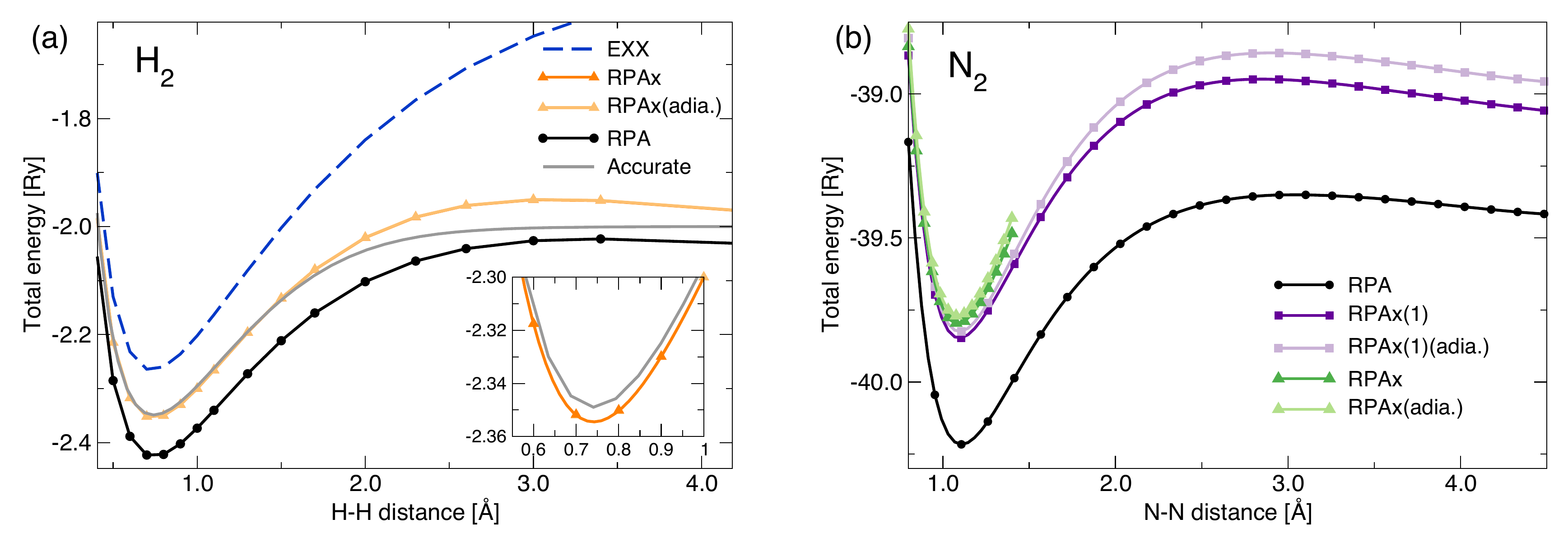}
\caption{(a) Dissociation energy curves of the hydrogen molecule in EXX (blue dashed lines), RPA (black circles) and RPAx ((light) orange triangles). Accurate result (gray) is reproduced from W. Kolos and L. Wolniewicz, J. Chem. Phys. {\bf 43}, 2429 (1965), with the permission of AIP Publishing. The inset shows a zoom around the bond midpoint. (b) Dissociation energy curves of the nitrogen molecule in RPA (black circles), RPAx ((light) green triangles), and with the alternative resummation RPAx(1) ((light) purple squares).}
\label{h2n2}
\end{figure*}
In this way, approximate interaction energies are generated via approximate xc kernels. 

We will now rewrite also the kinetic energy in terms of the xc kernel using a fixed-density adiabatic-connection (AC) together with Eq. (\ref{uxc}). The AC is defined by a parameter $\l$ that scales the electron-electron interaction linearly between 0 and 1, while keeping the density fixed to its $\l=1$ value via an added fictitious potential. At $\l=0$ the Hamiltonian is thus equal to that of the KS system and at $\l=1$ to that of the fully interacting system. 
By differentiating and then integrating the $\lambda$-dependent total energy it is possible to decompose the total energy at $\l=1$ as 
\be
E=T_s+E_{\rm ext}+U_0+E_{\xc},
\label{KSenergy}
\ee
where $T_s$ is the kinetic energy of KS electrons and the remaining xc part $T_\xc=T-T_s$ is incorporated in a new definition of the xc energy
\bea
E_{\xc}&=&-\frac{1}{2}\sum_{\s\s'}\int_0^1\!d\l\int \!d\vr d\vr'\,v(\vr-\vr')\nn
&&\!\!\!\!\!\!\!\!\!\!\!\!\!\!\!\!\!\!\!\!\!\!\times\left[\int_0^\infty \!\frac{d\w}{\pi} \chi^{\s\s'}_\l(\vr,\vr',i\w)+n_\s(\vr)\d_{\s\s'}\d(\vr-\vr')\right].
\label{exclamb}
\eea
which differs from $U_\xc$ only in the integration over $\l$, and coincides with the KS DFT definition of the xc energy. This is known as the ACFD xc energy. It allows us to determine the xc kinetic energy given an approximation for the $\l$-scaled $f_\xc$ from
\be
T_\xc=E_{\xc}-U_\xc.
\ee
 
We will now focus on the TDEXX approximation, which can be viewed as the TDDFT variant of TDHF. 
The nonlocal HF potential is replaced by the time-dependent local EXX potential $v_\x$, from
which the linear response equations are generated (Eqs.~(\ref{tddftLR}-\ref{fxceq})). 
The spin-resolved EXX kernel $f_\x$ is given by
\bea
&&\int \!d\vr_1d\vr_2\,\chi_s^{\s\s}(\vr,\vr_1,\w)f^{\s\s}_{\x}(\vr_1,\vr_2,\w)\chi_s^{\s\s}(\vr_2,\vr',\w)\nn
&&\,\,\,\,\,\,\,\,\,\,\,=R^{\s\s}_V(\vr,\vr',\w)+R^{\s\s}_\S(\vr,\vr',\w)
\label{fxeq}
\eea
and
\be
f^{\downarrow\uparrow}_{\x}(\vr,\vr',\w)=f^{\uparrow\downarrow}_{\x}(\vr,\vr',\w)=0.
\ee
The diagrammatic structure of this equation, as well as the analytic expression 
for the vertex $R_V$ and self-energy $R_\S$ terms can be found in e.g. Refs.~\onlinecite{PhysRevB.78.115107,hellgren_vonbarth2010,PhysRevResearch.3.033263}. The EXX kernel carries a frequency dependence due to the fact that $v_\x$ is nonlocal in time. The frequency dependence is well-defined and necessary to exactly reproduce the TDHF density to first order in $v$.\cite{PhysRevB.78.115107} It is, however, possible to apply the adiabatic approximation also within TDEXX by evaluating the EXX kernel at $\w=0$ and then using this kernel at every frequency in Eq. (\ref{tddftLR})
\be
 f_\x\approx  f_\x(\w=0).
\ee
The use of the adiabatic EXX (AEXX) kernel has the potential to reduce the computational 
cost of the calculations since Eq.~(\ref{fxeq}) only needs to be evaluated once, instead of around 8 times as in a typical calculation on a gapped system. In the following we will investigate the effect of this approximation for calculating correlation energies. 

As in previous notation, the total energy that uses the fully frequency dependent EXX kernel in conjunction with Eq.~(\ref{KSenergy}) will be denoted RPAx. The approximation generated by replacing the EXX kernel in RPAx with the AEXX kernel will be denoted RPAx(adia.). Ignoring the xc kernel completely gives the standard total energy within RPA.

\section{Numerical study}
In this section we give the computational details and present a numerical study on the stretched H$_2$ and N$_2$ molecules, the strongly correlated H-chain (equidistant and dimerized) and the weakly bonded H$_2$-dimer, solid-Ar and the water dimer. Finally, in order to carefully investigate how the adiabatic approximation affects both the correlation energy and the spectrum we present a study of a model system of dissociating closed-shell atoms.

\subsection{Computational details}
All calculations have been performed with an updated version of the ACFDT code within the Quantum ESPRESSO package.\cite{Giannozzi_2017,PhysRevB.90.125150,PhysRevB.93.195108,hellgren_correlation_2018,PhysRevResearch.3.033263} The RPAx correlation energy is determined by first solving a 
generalized eigenvalue problem based on Eq.~(\ref{fxeq}) 
\be
 R|u_\b\ket=\nu_\b\chi_s|u_\b\ket,
\label{eig}
\ee
where $R=R_V+R_{\S}+\chi_s v\chi_s$ is expressed in the eigenpotentials of $\chi_s$ obtained in an initial step by iterative diagonalization.\cite{PhysRevB.79.205114} The set of eigenvalues $\nu_\b$ and eigenvectors $|u_\b\ket$ are generated at every frequency and $\bf q$-point in the Brillouin zone. The correlation energy is then computed from the following expression 
\bea
E_{\rm c}&=&-\frac{1}{N_{\bf q}}\sum_{\bf q}\sum^{N_\nu}_\b\int_0^\infty \frac{d\w}{2\pi}\frac{\bra u_\b|\chi_s v\chi_s|u_\b \ket}{\nu_\b({\bf q},i\w)}\nn
&&\,\,\,\,\,\,\,\,\,\,\,\,\,\,\,\,\,\,\,\times\{\nu_\b({\bf q},i\w)+\ln[1-\nu_\b({\bf q},i\w)]\},
\label{echi}
\eea
where $N_\nu$ is the total number of eigenvalues and $N_{\bf q}$ is the total number of $\bf q$-points. The number of eigenvalues needed for convergence of energy differences is typically around 10 times the number of electrons. More convergence analyses can be found in Refs. \onlinecite{hellgren_correlation_2018,PhysRevResearch.3.033263}. 

In the case of the adiabatic approximation we first determine $f_\x(\w=0)$ in the eigenpotentials of $\chi_s(\w=0)$. At finite frequency, Eq.~(\ref{eig}) is solved by computing $R$ from the left-hand side of Eq.~(\ref{fxeq}), in which $f_\x(\w)$ is approximated by $f_\x(\w=0)$ projected on the space spanned by the eigenvectors of $\chi_s(\w)$. 

All calculations have been done non-self-consistently on top of PBE orbitals. The electron-ion interaction is treated with standard norm-conserving pseudopotentials.\cite{PhysRevB.88.085117,PhysRevB.54.1703,SCHLIPF201536}
\subsection{H$_2$ and N$_2$ dissociation}
The H$_2$-dissociation curve has been analysed in many previous works.\cite{fuchs_describing_2005,hellgren_correlation_2012,PhysRevB.91.165110,colonna_correlation_2014,doi:10.1063/1.4871875,PhysRevLett.106.093001} In Fig.~\ref{h2n2}~(a) the results are reproduced in order to verify that in the case of two-electron systems the EXX kernel is frequency independent (and equal $-v/2$). RPAx and RPAx(adia.) are indistinguishable on the scale of the figure, differing by less than 0.0002 Ry. The difference gets smaller the more eigenvalues are included in the calculation. We note the high accuracy of the RPAx total energy around equilibrium geometry.\cite{doi:10.1063/1.1697142} However, thanks to a good cancellation of error the RPA binding energy is comparable and even slightly better than RPAx.\cite{colonna_correlation_2014}
\begin{figure*}[t]
\includegraphics[scale=0.56]{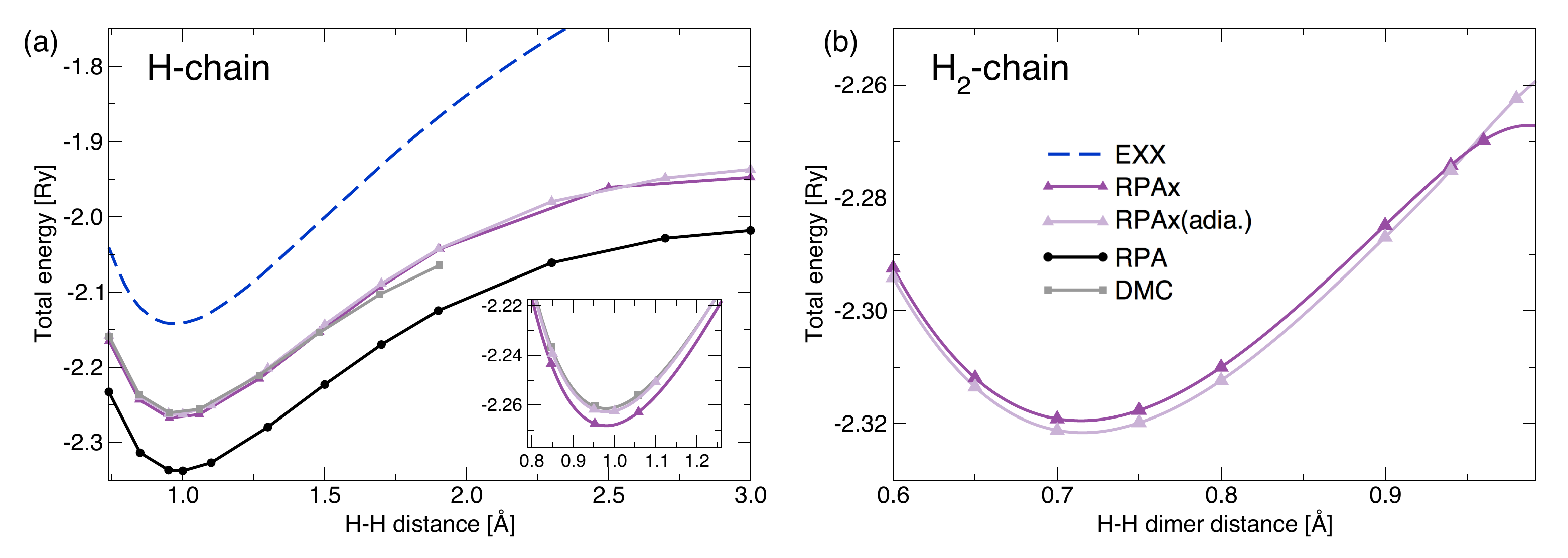}
\caption{(a) Dissociation energy curve of the equidistant hydrogen chain in EXX (blue dashed lines), RPA (black circles) and RPAx ((light) purple triangles). DMC results (gray) are from Motta et al., Phys. Rev. X {\bf 7}, 031059, (2017); licensed under a Creative Commons Attribution (CC BY) license. The inset shows a zoom around equilibrium. (b) Dimerisation of the hydrogen chain starting from the equidistant H-chain at 0.98 \AA.}
\label{hchain}
\end{figure*}

In Fig.~\ref{h2n2}~(b) we plot the N$_2$-dissociation curve. We have also added RPAx(1), obtained from an alternative re-summation of the RPAx Dyson's equation that keeps $f_\x$ to first order only.\cite{colonna_correlation_2014,hellgren_correlation_2018} 
The difference between RPAx and RPAx(adia.) is larger in this case. The adiabatic approximation leads to an underestimation of the correlation energy. However, relative to the total binding energy the effect is rather small and the equilibrium bond distance changes by 0.002 {\AA} only. 

As the bond is stretched the RPAx breaks down due to an instability in the TDEXX response function. Since the AEXX kernel gives rise to an identical instability we can conclude that the frequency dependence is irrelevant for this behaviour. As shown previously, using the RPAx(1) re-summation, such instabilities are circumvented without loss of accuracy.

\subsection{Strongly correlated H-chain}
Let us now look at a periodic system. In Fig.~\ref{hchain}~(a) we plot the total energy per atom as a function of atomic separation of the equidistant hydrogen chain. The effect of the AEXX kernel is again to slightly underestimate the correlation energy. The equilibrium geometry is, however, almost unaffected, staying at 0.98 {\AA}. We have also included recent DMC results\cite{PhysRevX.7.031059} in order to demonstrate the accuracy of the RPAx total energy around equilibrium. The error stays below 4 mRy/atom. 

In Fig.~\ref{hchain}~(b) we plot the energy gain upon dimerization, starting from the equidistant chain at 0.98 {\AA}. In this way, we form a chain of H$_2$ molecules with {1.96 \AA} between the centre of the molecules. Due to the interaction between the molecules 
there is a non-zero frequency dependence of the EXX kernel. Indeed, RPAx(adia.) differ from RPAx, in this case by overestimating the correlation energy. However, similarly to the cases above the 
energy gain and position of the minimum are only weakly affected. 
\begin{figure}[b]
\includegraphics[scale=0.33]{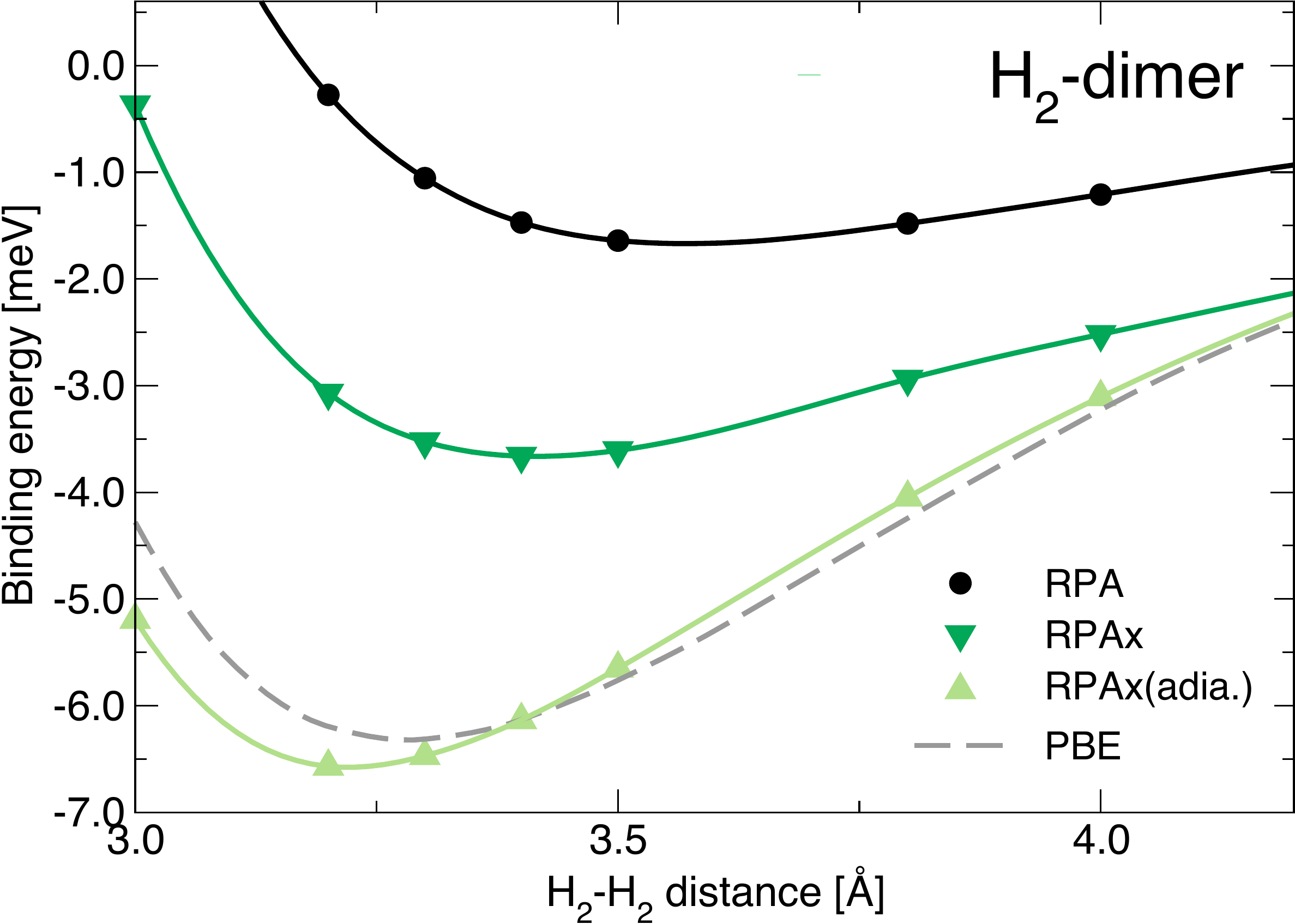}
\caption{Binding energy curve of the H$_2$-H$_2$ dimer in the T-configuration. RPA (black circles), RPAx ((light) green triangles) and PBE (gray dashed line).}
\label{h2h2}
\end{figure}

\subsection{Non-covalent bonds}
The interaction between the molecules in the H$_2$ chain is mainly repulsive but could have a van der Waals minimum at larger separation. Such a weak bond ($\approx$ 1 meV) is found in an isolated dimer of H$_2$ molecules when aligned linearly.\cite{diep2000accurate,h2dimers} A more stable configuration is found in the T-shaped dimer (36.2 cm$^{-1}$ in CCSD(T)\cite{h2dimers}). In Fig.~\ref{h2h2} we plot the binding energy curve of such a T-dimer. We compare PBE, RPA, RPAx and RPAx(adia.). The RPA is seen to underestimate the binding energy while RPAx provides an improvement. We also see that there is a relevant difference between RPAx and RPAx(adia.). The latter produces a binding energy around twice as large and 
underestimates the bond distance (3.405 {\AA} in CCSD(T)\cite{h2dimers}). Since H$_2$ is a two-electron system the energy in the dissociation limit should agree. This means that the difference in total energy between RPAx and RPAx(adia.) is equal to the difference in binding energy ($\approx$ 2 meV around equilibrium). Such difference can be considered small in the case of a strong covalent bond but leads to significant discrepancies for describing a weak bond. 

\begin{figure}[t]
\includegraphics[scale=0.5]{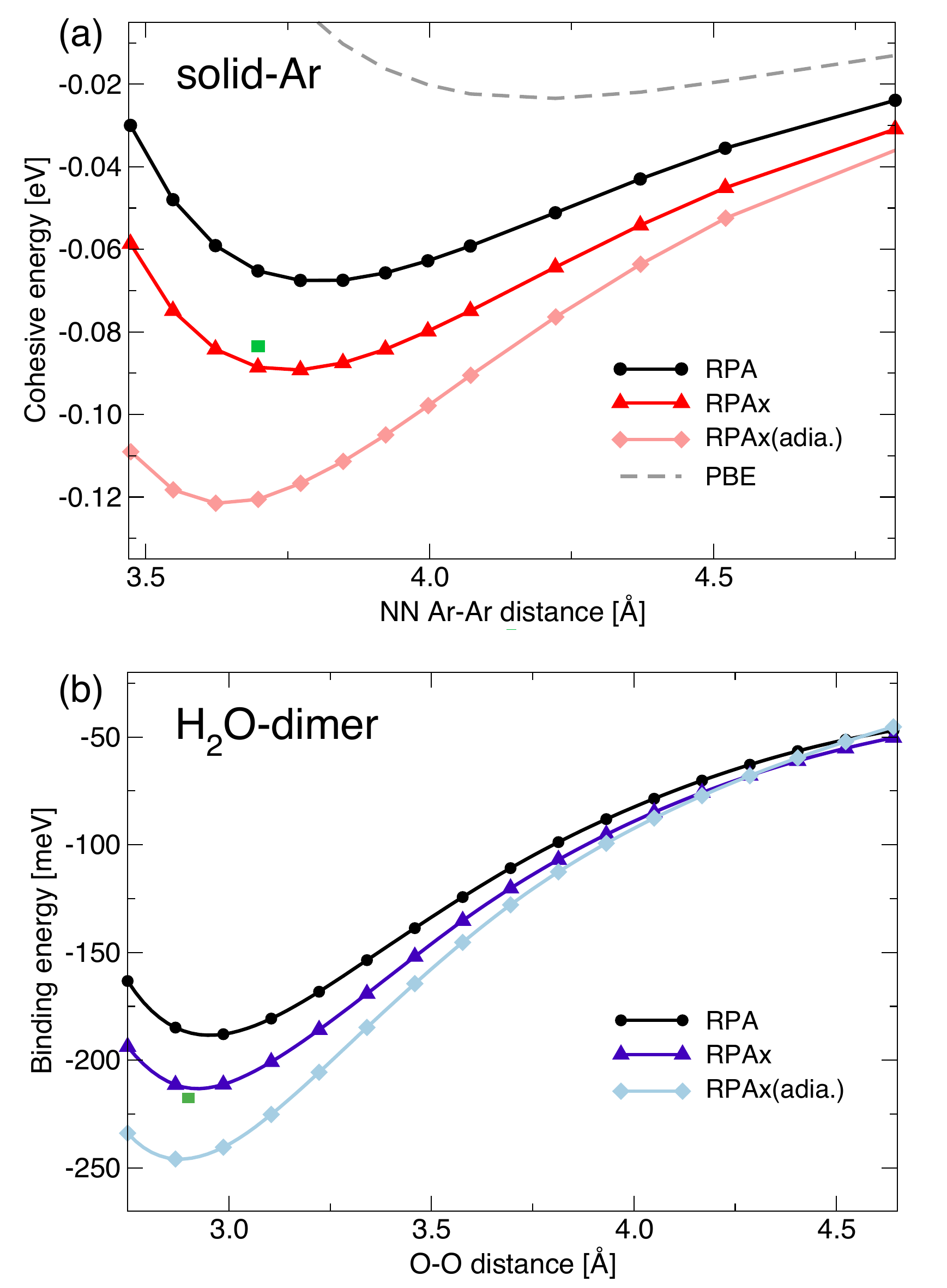}
\caption{(a) Cohesive energy of solid Argon as a function of nearest neighbour (NN) distance. PBE (gray dashed lines), RPA (black circles) and RPAx ((light) red diamonds). Green square corresponds to the CCSD(T) result. Reproduced with permission from Phys. Rev. B {\bf 60}, 7905 (1999). Copyright 1999 American Physical Society. (b) Binding energy of the water dimer in the lowest energy hydrogen bond configuration. RPA (black circles) and RPAx ((light) blue triangles). Green square corresponds to the CCSD(T) result.\cite{doi:10.1063/1.1408302} Reproduced from G. S. Tschumper et al., J. Chem. Phys. {\bf 116}, 690 (2002), with the permission of AIP Publishing.}
\label{arice}
\end{figure}

To see whether the overestimation of the binding energy produced by the AEXX kernel is a general trend for non-covalent systems we have also studied solid-Ar (van der Waals bond) and the H$_2$O dimer (mixed van der Waals and hydrogen bond). In Fig.~\ref{arice}~(a) we plot the cohesive energy of solid-Ar as a function of nearest neighbour distance. Again, 
the AEXX kernel overestimates the interaction energy. The absolute error compared to CCSD(T)\cite{PhysRevB.60.7905} is almost twice as large as when comparing RPA to CCSD(T). The effect is somewhat smaller but qualitatively the same in the H$_2$O dimer as demonstrated in Fig.~\ref{arice}~(b). 

It has been shown that using the adiabatic approximation for calculating van der Waals (or $C_6$) coefficients has 
a negligible effect.\cite{PhysRevA.73.010502,hellgren_vonbarth2010} However, the $C_6$-coefficient only describes 
the van der Waals interaction asymptotically. The error seen here gets larger the closer the atoms are and is thus of a different origin.  

\begin{figure*}[t]
\includegraphics[scale=0.5]{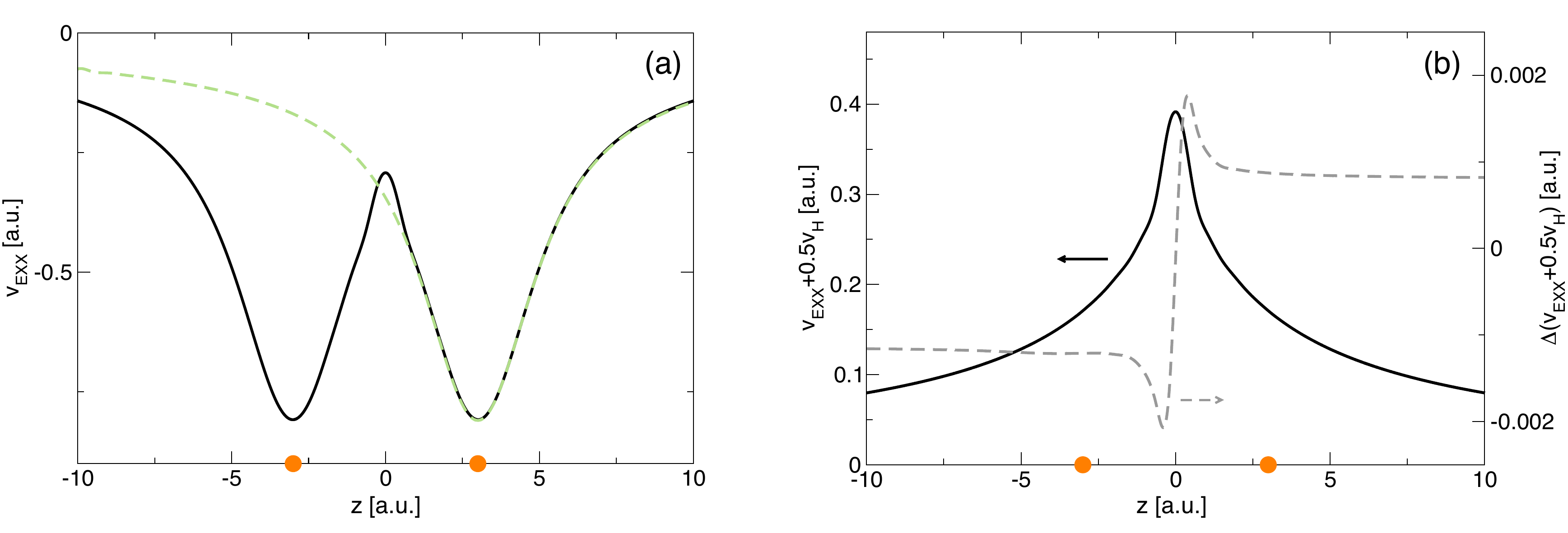}
\caption{(a) The EXX potentials of the 1D dimer at $R=6$ a.u. (black line) and the two-electron 1D atom (green dashed line). The position of the atoms are marked in orange. (b) The peak of the potential in (a) is exhibited by adding the Hartree potential divided by two (black line with y-axis to the left). The field-counteracting step that arises in response to an external linear potential (here $-0.001x$) is also shown (grey dashed line with y-axis to the right). }
\label{1dpot}
\end{figure*}

\begin{figure}[b]
\includegraphics[scale=0.5]{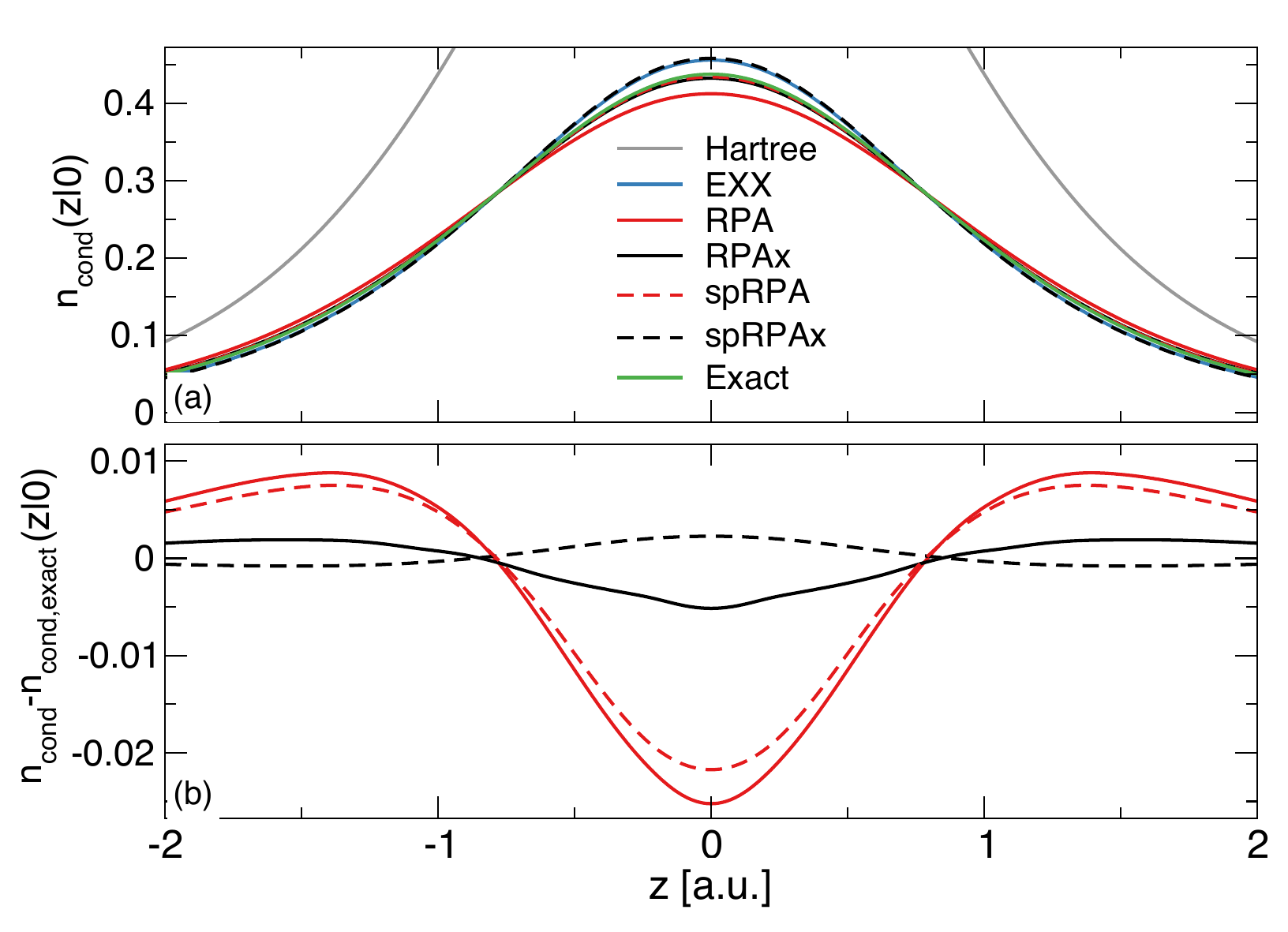}
\caption{(a) The conditional probability density of the two-electron model 1D atom in Hartree, EXX, RPA, RPAx and the exact calculation. Including only the same-spin component (sp) within RPA/RPAx is shown with dashed lines. The self-consistent RPAx potential is used to generate the KS orbitals. (b) The difference with respect to the exact calculation.}
\label{1dcond}
\end{figure}

\subsection{1D dimer}
In order to investigate the origin of the difference in correlation energy due to the adiabatic approximation in non-covalent systems we will here study a model one-dimensional (1D) dimer composed of two-electron atoms at various separation $R$ (e.g. a model He-He or H$_2$-H$_2$ dimer). The electron-electron interaction 
\be
v(z-z')=\frac{1}{\sqrt{(z-z')^2+1}}
\ee
and external potential 
\be
v_{\rm ext}(z)=-\frac{2}{\sqrt{(z-R/2)^2+1}}-\frac{2}{\sqrt{(z+R/2)^2+1}}
\ee
are given along the $z$-axis and described by soft Coulomb interactions. The KS and TDEXX equations are solved by expanding the KS orbitals in a cubic spline basis set. More details regarding the numerical method can be found in Refs. \onlinecite{PhysRevB.76.075107,PhysRevA.85.022514,PhysRevA.88.052507}. The exact numerical solution of the Schr\"odinger equation for a 1D two-electron atom is obtained using the OCTOPUS code,\cite{2006octopus} via a mapping of the two-electron 1D system to a one-electron 2D system.

Let us start by looking at the ground state within the EXX approximation. The EXX potential at $R=6$ a.u. is plotted in Fig.~\ref{1dpot}~(a), superimposed on the atomic 
two-electron EXX potential (green dashed line). The largest effect of the presence of a second atom is the peak 
located between the atoms. To single out this peak feature, the Hartree potential divided by two 
(which is exactly equal to the negative of the EXX potential in the dissociation limit) 
is added to the EXX potential in Fig.~\ref{1dpot}~(b). The peak is now clearly visible and can be seen as analogous to the correlation-peak that has been extensively studied in the study of bond dissociation into open-shell atoms (e.g. in H$_2$ dissociation). The correlation-peak is often observed together with a step feature when looking at heteronuclear systems such as LiH.\cite{PhysRevA.54.1957,tempel2009revisiting,hellgren2019strong} To show that such a step is present also in the EXX potential we apply a weak electric field that breaks the symmetry of the system. 
The response of the EXX potential is shown in Fig.~\ref{1dpot}~(b) (grey dashed curve) and, indeed, a step feature appears. A similar effect was studied in H$_2$ chains where it is known as the field counteracting effect, important to describe polarizabilities as it prevents electron delocalization over the chain.\cite{gisbergen,kummelchain} 

At the static ground-state level we thus see an important role of the exchange potential in our model. Features such as peaks and steps, not present in semi-local functionals, act to localize charges on the atoms, playing a similar role as they do in the correlation potential of dimers composed of open-shell atoms. 

The response step, as demonstrated in Fig.~\ref{1dpot}~(b), was, in Ref.~\onlinecite{PhysRevA.88.052507}, given a perspective in terms of diverging steps in the EXX kernel. It was shown that steps in the kernel that grow larger with atomic separation are essential to compensate for the sometimes vanishing KS excitation functions in linear response TDDFT. The step-shape ensures that local properties are not affected. Furthermore, the frequency-dependence was shown to strongly enhance the magnitude of the step, enabling a correct description of e.g. charge-transfer excitations.\cite{PhysRevA.85.022514} 

Let us now investigate how the frequency dependence of the EXX kernel influence the correlation energy obtained from Eq.~(\ref{uxc}). First of all, as mentioned above, in the special case of two-electron systems the EXX kernel coincides with the AEXX kernel and is therefore frequency independent. The RPAx total energy of the isolated (or dissociated) two-electron atom is thus identical to the RPAx(adia.) total energy. Comparing with our exact wave-function calculation we find that the RPAx total energy is 
very close to the exact total energy (-2.2379 a.u. and -2.2382 a.u. respectively), which can be compared to the RPA total energy of -2.2486 a.u.. These energies are obtained self-consistently, i.e., the RPA/RPAx potential is also calculated. 

More insights can be gained by looking at the diagonal of the two-particle density matrix, Eq.~(\ref{defg2}), accessible both from $\chi$, via 
$f_\xc$, and the wave function of the exact calculation. In Fig.~\ref{1dcond}~(a) we plot the conditional probability, Eq.~(\ref{ncond}), for finding an electron at $\vr$ given another one is at the center of the model atom. We compare the exact calculation (spin-summed), the Hartree (or independent) approximation (first term of Eq.~(\ref{ncond})), EXX, RPA and RPAx (fully spin-summed and by including only the same-spin component (sp) in RPA/RPAx). In this system EXX is exactly half of Hartree as exchange introduces a same-spin correlation that removes the probability of finding an electron of the same spin anywhere in the system. This is an exact result that should be preserved by the same-spin component in RPA/RPAx. We see that RPAx correctly remains almost on top of EXX while RPA generates an unphysical negative probability of finding an electron of the same spin at the same point (see also Fig.~\ref{1dcond}~(b)). By looking at the fully spin-summed conditional probability in RPA and RPAx we see that the error in RPA is mostly in the same-spin component. The RPAx is in very good agreement with the exact wave function result.
\begin{figure}[b]
\includegraphics[scale=0.33]{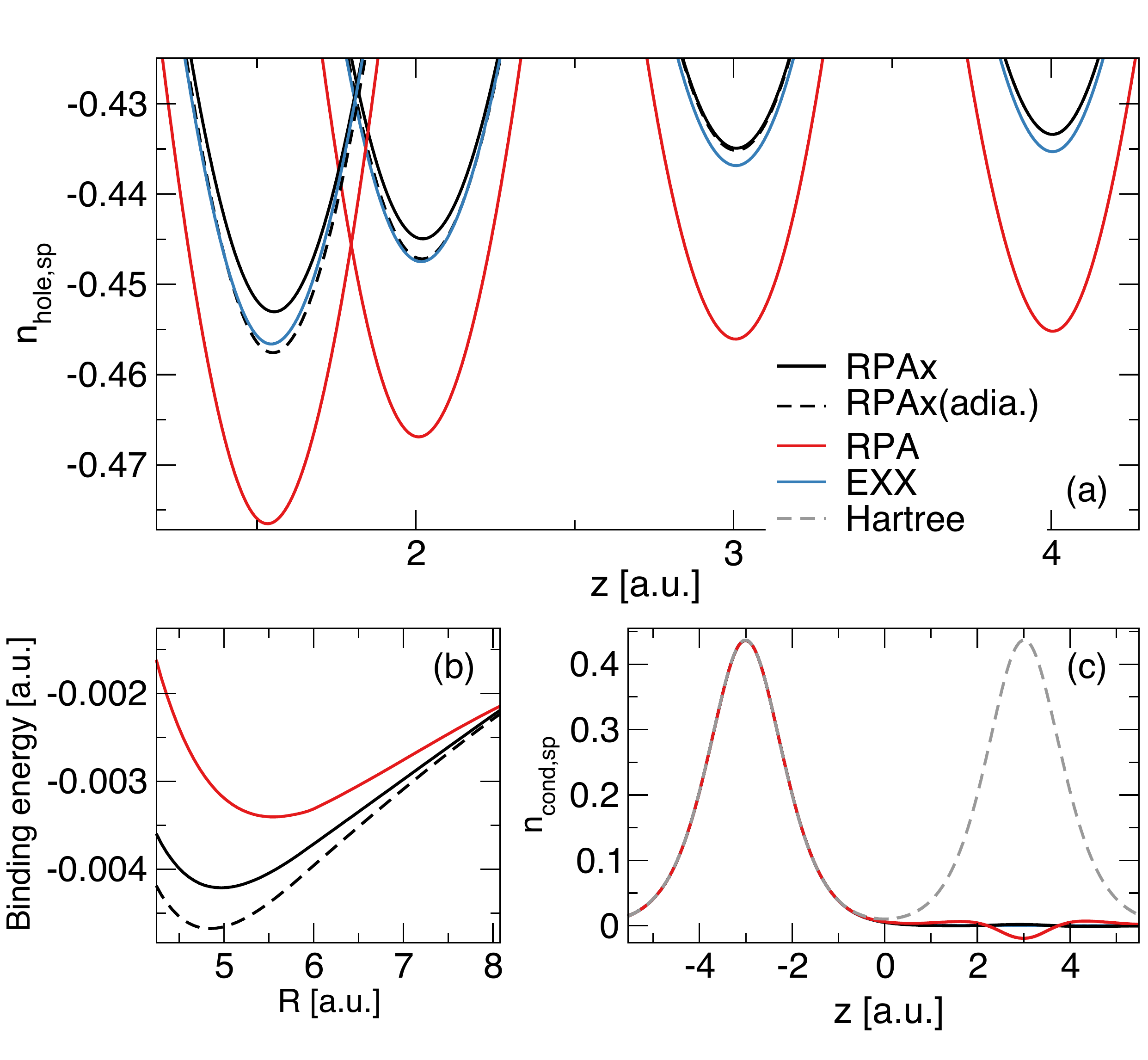}
\caption{(a) The same-spin xc hole around one of the atoms of the dimer at various atomic separations ($3,4,6$ and 8 a.u.). (b) Binding energy as a function of atomic separation. (c) The full  same-spin $n_{\rm cond}$ in Hartree, EXX, RPA, RPAx and RPAx(adia.). }
\label{1ddiss}
\end{figure}
\begin{figure*}[t]
\includegraphics[scale=0.55]{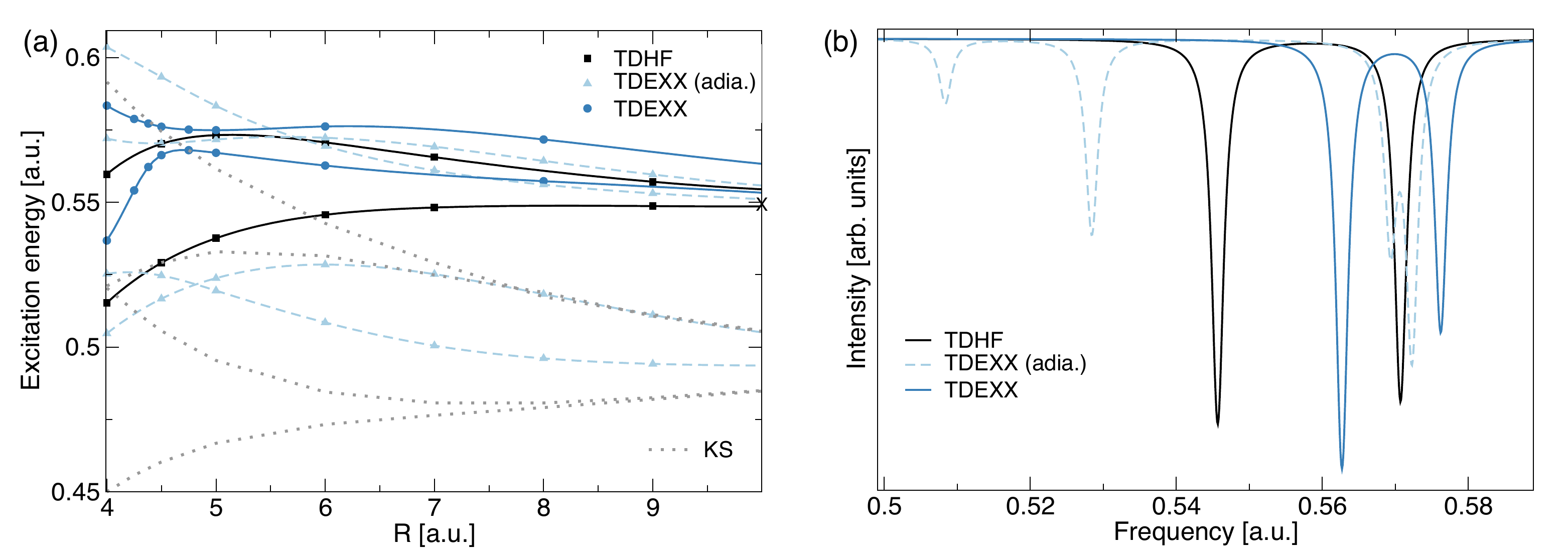}
\caption{(a) Excitation energies as a function of atomic separation of the 1D model dimer in KS, TDEXX, TDEXX(adia.) and TDHF. (b) Spectrum at $R=6$ a.u. in TDEXX, TDEXX(adia.) and TDHF.}
\label{excit}
\end{figure*}

We now turn to the combined system of two two-electron atoms. In this four-electron system the EXX kernel carries a frequency dependence. Only in the dissociation limit will the RPAx and RPAx(adia.) total energies coincide. In Fig.~\ref{1ddiss}~(a) we plot the evolution ($R=3,4,6,8$ a.u.) of the same-spin xc hole (Eq.~(\ref{ncond})) over one of the atoms and, indeed, at $R=8$ a.u. the RPAx and RPAx(adia.) xc holes are visually indistinguishable. The small overestimation of the spRPAx conditional probability in Fig.~\ref{1dcond} is here seen as a small underestimation of the depth of the same-spin xc hole. 
As the atoms gets closer the depth remains slightly underestimated in RPAx, while in RPAx(adia.) the hole gets deeper and eventually even slightly deeper than in EXX, implying negative probabilities for same-spin electrons. An increased probability of having two electrons at the same point increases the interaction energy. Since the differences between RPAx and RPAx(adia.) is almost entirely in the same-spin component the correlation energy is expected to be larger within RPAx(adia.). 
Indeed, in Fig.~\ref{1ddiss}~(b) we show how the interaction energy of the dimer is overestimated with the AEXX kernel, a behaviour which is very similar to the behaviour of the real noncovalent dimers studied in the previous section. Although the on-top RPAx(adia.) same-spin xc hole appears to be in overall better agreement with the exact EXX result of $-n_\s(\vr)$ at intermediate separation the evolution is less consistent. This suggests that the improvement is for the wrong reason and the error in describing the dimer do not cancel when taking energy differences. This explains why the binding energy is worse. The results for reaction energies of small molecules in Ref.~\onlinecite{doi:10.1080/00268970903476662} point to a similar problem.

Finally, in Fig.~\ref{1ddiss}~(c) we plot the same-spin conditional probability of the full dimer. We see how the EXX approximation cancels the Hartree probability density around the reference atom, eliminating the probability of finding another electron with the same spin there. We also see how RPA predicts a large unphysical negative probability.

The correlation energies and probability densities are calculated by performing the frequency integral in Eq.~(\ref{uxc}) on the imaginary axis where $\chi$ is smooth. In order to analyse the effects of the 
adiabatic kernel on the physical properties of $\chi$ itself we will now look at the excitation spectrum 
obtained from the pole structure on the real frequency axis. This will give us a 
second perspective of why the adiabatic approximation is problematic on this kind of systems. 

At intermediate to large separation the dimer has two occupied KS orbitals 
formed by the symmetric and antisymmetric combination of the occupied atomic orbitals. 
Similarly, the first two unoccupied KS orbitals correspond to the symmetric and antisymmetric combination of the first unoccupied atomic orbitals. We will now restrict ourselves to the excitations coming from these orbitals and study the spectrum with and without the frequency dependence of the kernel.

In Fig.~\ref{excit} (a) we show the first four excitation energies as a function of atomic separation and in Fig.~\ref{excit} (b) the corresponding spectrum at 6 a.u.. The latter is obtained by taking the imaginary part of $\chi$ using a broadening of $\d=0.001$ a.u..
As expected, all KS (EXX) excitation energies approach the same atomic KS eigenvalue difference in the dissociation limit. A qualitatively different behaviour is found within TDEXX and TDEXX with the AEXX kernel (TDEXX(adia.)). Only two of the excitations reach the first atomic TDEXX excitation in the dissociation limit (marked with a black cross and the same with or without frequency dependence). In TDEXX(adia.) two CT excitations are identified at lower energy. The CT excitation energy is underestimated and eventually coincide the KS excitations, i.e. the correction due to the Coulomb and AEXX kernel goes to zero with separation. In TDEXX, with the frequency dependence taken into account, the CT excitations disappear from the spectrum already at intermediate separation and most of the weight is transferred to the local excitations. 

To have a reference we also performed TDHF calculations on the same system. Again two excitations 
corresponding to the local atomic excitations are identified with a magnitude similar to TDEXX.
The CT excitations are correctly transferred to much higher energies where they have a small but non-zero weight. Overall the behaviour is more similar to TDEXX, illustrating that the frequency dependence of the kernel tries to mimic certain important features of the TDHF spectrum. The use of the adiabatic kernel within TDEXX leads to a large underestimation of the CT excitation energies and an overestimation of their weight at intermediate atomic separation.  

\section{Conclusions} \label{SEC:conclusions}
The RPAx within the ACFD framework has been shown to produce accurate correlation energies 
on a wide range of molecular and solid-state systems. In this work we have investigated the impact of using the adiabatic 
approximation to the EXX kernel. The main motivation was to investigate the possibility of reducing 
the computational cost, but also to gain a deeper theoretical insight into the role of the frequency dependence 
of the kernel. 

Starting with a study on H$_2$, N$_2$ and the equidistant and dimerized H-chain we confirmed previous 
works that have suggested that the adiabatic 
approximation is sufficient for calculating correlation energies. However, when studying non-covalent
systems, such as the H$_2$-dimer, solid-Ar and the H$_2$O-dimer, significant errors around equilibrium 
geometry were found. The AEXX kernel overestimates the interaction energy which leads to shorter 
bond lengths and too large binding energies. 

Since the failure of the AEXX kernel for non-covalent interactions cannot be attributed to a different 
C$_6$-coefficient, we carried out a more detailed study on a simplified 1D model of two closed-shell atoms at 
various separation. 
The evolution of the xc hole was studied within RPAx and RPAx(adia.) demonstrating that the AEXX kernel produces 
a deeper same-spin xc hole at intermediate separation and eventually negative probabilities that do not cancel when looking at energy differences. 

Finally, another perspective of why RPAx(adia.) 
fails in these kind of systems was provided by looking at the spectrum obtained from $\chi$. CT excitations 
in the low energy part of the spectrum turn out to be responsible for a strong frequency dependence in the 
EXX kernel, necessary to correctly describe both the weight and magnitude of the excitation energies.

In conclusion, our results for the EXX kernel show that the frequency 
dependent kernel cannot, in general, be replaced by its adiabatic approximation 
for ground-state total energy calculations. Since the EXX kernel has a well-defined frequency dependence, 
our conclusions are expected to carry over to more advanced kernels.

\begin{acknowledgments}
The work was performed using HPC resources from GENCI-TGCC/CINES/IDRIS (Grant No. A0110907625). 
Financial support from Emergence-Ville de Paris is acknowledged.
\end{acknowledgments}

%

\end{document}